\documentclass[twocolumn,reprint]{revtex4}

\usepackage{graphicx}
\usepackage{dcolumn}
\usepackage{bm}

\begin{document}


\title{Low frequency noise characteristics of sub-micron magnetic tunnel junctions} 



\author{B. Zhong}
 \email{zhong@physics.sc.edu}
\author{Y. Chen}
 \altaffiliation{Now at Department of Physics \& Astronomy, Rutgers University, Piscataway, NJ 08854}
\author{S. Garzon}
 \altaffiliation{Now at San Jose Research Center, Hitachi Global Storage Technologies, San Jose, CA 95120}
\author{T.M. Crawford}
\author{R.A. Webb}
 \affiliation{Department of Physics \& Astronomy and USC Nanocenter, University of South Carolina, Columbia, SC 29208\\}


\date{\today}

\begin{abstract}
We report that low frequency (up to 200 kHz) noise spectra of magnetic tunnel junctions with areas ~10$^{-10}$cm$^2$ at 10 Kelvin deviate significantly from the typical $1/f$ behavior found in large area junctions at room temperature. In most cases, a Lorentzian-like shape with characteristic time between 0.1 and 10 ms is observed, which indicates only a small number of fluctuators contribute to the measured noise. By investigating the dependence of noise on both the magnitude and orientation of an applied magnetic field, we find that magnetization fluctuations in both free and reference layers are the main sources of noise in these devices. At small fields, where the noise from the free layer is dominant, a linear relation between the measured noise and angular magnetoresistance susceptibility can be established.
\end{abstract}


\maketitle 

\section{INTRODUCTION}

Magnetic tunnel junctions (MTJs) are currently used for information readout in disk drives~\cite{wood_JMMM2009}, and for high sensitivity magnetic field sensors~\cite{egelhoff_SAA2009}, and could be used in future technologies such as spin random access memory~\cite{katine_JMMM2008}. For all these applications, the noise generated at the MTJs is a crucial factor limiting their ultimate performance. Besides the ubiquitous frequency independent thermal and shot noise present in MTJs, frequency dependent noise sources such as charge trapping in the oxide barrier~\cite{nowak_APL1999} and thermally excited magnetization fluctuations and domain wall oscillations~\cite{Ingvarsson_PRL2000} have been identified. For MTJs with areas larger than a few square microns, these noise sources usually exhibit a $1/f$ frequency dependence due to ensemble averaging of many individual fluctuators~\cite{liu_JAP2003,Jiang_PRB2004,Ren_PRB2004,nor_JAP2006,guo_APL2009}. In that case the magnitude of the $1/f$ noise is related to the resistance susceptibility as predicted by the fluctuation-dissipation theorem which only applies in thermal equilibrium~\cite{Ingvarsson_PRL2000}. However, as the dimensions of MTJs are reduced to the single domain range, the applicability of ensemble averaging and the fluctuation-dissipation relation are both under question.

In this work, we report on measurements of low frequency noise (up to 200 kHz) of single domain MTJs with areas $\sim$10$^{-10}$cm$^2$ at a temperature of 10 Kelvin. Our results show that for MTJs of this size, the dominant source of low frequency noise are magnetization fluctuations of both free and reference layers. For magnetic fields, $B$, well below the pinning field of the reference layer, $B_{AF}$, noise from the free layer dominates, and a linear relation between noise magnitude and angular magnetoresistance (MR) susceptibility, $\chi_{MR}=(dR/d\theta)/R$ (the derivative of magnetoresistance with respect to magnetic field orientation $\theta$), is observed. As $B$ increases and becomes comparable to $B_{AF}$, noise originating from magnetic fluctuations of the reference and free layer become comparable, and the linear relation between noise and $\chi_{MR}$ no longer holds.

\section{EXPERIMENT}
\begin{figure}
\begin{center}
\includegraphics[width=3.4in]{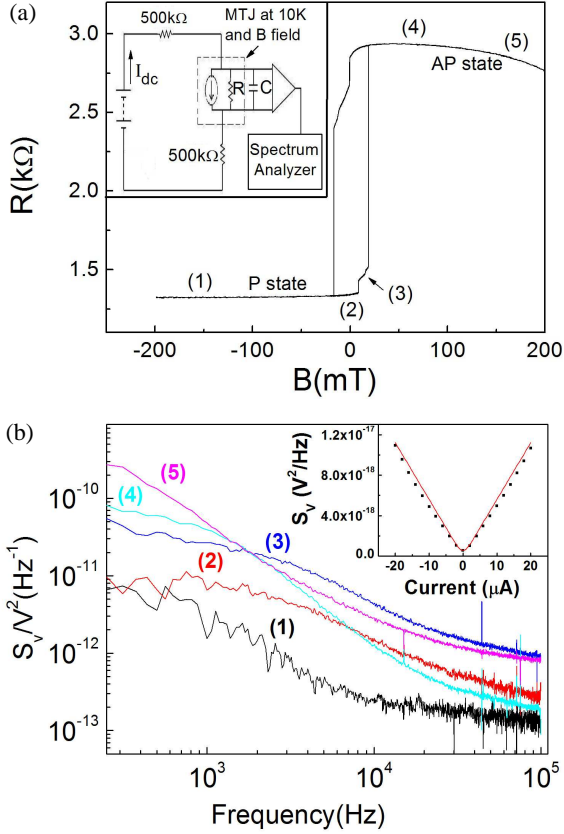}\\
\caption{(a) Resistance vs. easy axis field at 5 $\mu$A and 10 K. The resistance decrease at $B>80$ mT results from destabilization in the reference layer. Inset: experimental setup and equivalent noise model, showing two 500 k$\Omega$ isolation/bias resistors, an EG$\&$G 5186 differential voltage preamplifier, and an Agilent 89410A Vector Signal Analyzer. (b) Typical noise spectra at each of the regions labeled in hysteresis loop. Inset: average noise (black squares) in the 70-90 kHz frequency band at region 1 as a function of current; the red solid line represents Eq.~\ref{eq:shotnoise}, without any free fitting parameters.}\label{fig:loops}
\end{center}
\end{figure}

We use MTJs with the following stack composition (in $\AA$): buffer-layer/PtMn(175)/synthetic-antiferromagnet-
pinned-layer (SAF)/MgO(9.5) /Fe(5) /Co$_{60}$Fe$_{20}$B${_20}$(10) /
Ru(100)/capping-layer (see Ref. ~\cite{guan_JAP2009} for fabrication details). Devices are patterned into nanopillars with dimensions $\sim$100$\times$150 nm, where the easy axis of the free layer is nominally collinear with the pinning direction of the reference layer. The devices are cooled down to 10 K and can be rotated to any angle with respect to an applied in-plane magnetic field. The setup for the noise measurements is shown in the inset to Fig.~\ref{fig:loops} (a). The system has been calibrated using metal film resistors as sources of thermal noise and we have observed an almost flat spectrum with small roll-off that can be accurately modeled by a stray capacitance of $\sim$360 pF due to the cryostat wiring and limited amplifier bandwidth. The resolution of this setup is better than 0.3 nV/Hz$^{1/2}$ with one minute of averaging. Background measurements at 0 dc current are always subtracted from the raw data. The data presented here is for a single device, but similar results were obtained for two devices.

\section{RESULTS AND DISCUSSION}

\subsection{Noise as a function of magnetic field}
Fig.~\ref{fig:loops} (a) shows the hysteresis loop along the easy axis using $I_{dc}=5$ $\mu A$ (flowing from the free to the reference layer). Typical spectra at each of the labeled regions of Fig.~\ref{fig:loops} (a) are shown in Fig.~\ref{fig:loops} (b). We observe that the spectra do not agree with $1/f$ behavior, but can be described by a Lorentzian behavior with characteristic timescales from 0.1 to 10 ms, indicating that only a small number of fluctuators contribute to the measured noise~\cite{Dutta_PRL1979}. At large negative fields (region 1), where the field is parallel to the magnetic moments of both the free and the reference layers, the spectrum becomes almost flat above 70 kHz. The average noise between 70 and 90 kHz as a function of current is shown in the inset of Fig.~\ref{fig:loops} (b), and is close to the expected thermal and shot noise given by
\begin{equation}
S_V=2 e I_{dc} R^{2}\coth \left( \frac{e I_{dc} R}{2 k_B T} \right),
\label{eq:shotnoise}
\end{equation}

\noindent where R is the device resistance and $k_B T$ is the thermal energy. As B approaches zero, (region 2) and the free layer becomes bistable, the noise gradually increases up to the point where a sudden jump to an intermediate resistance state occurs, accompanied by an order of magnitude increase in noise (region 3). Once the free layer completely switches to the antiparallel (AP) state (region 4), the total noise power decreases and the Lorenzian corner frequency shifts towards lower frequency. At large positive fields (region 5), the resistance decreases since $B$ becomes comparable to $B_{AF}$, and starts to depin the layers within the SAF. The noise shows a more $1/f$-like behavior, indicating that the decoupling of the SAF layer produces a large number of uncorrelated fluctuations with different energy scales~\cite{Dutta_PRL1979}.

\subsection{Noise as a function of magnetic field orientation}
\begin{figure}
\begin{center}
\includegraphics[width=3.5in]{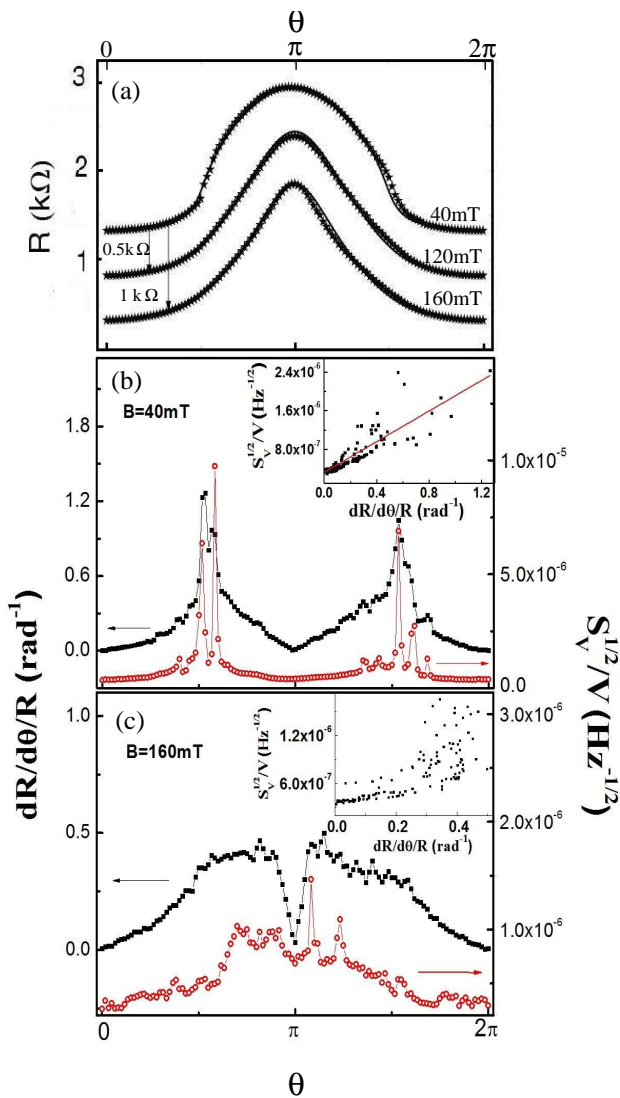}\\
\caption{(a) Resistance vs. magnetic field orientation for $B$ = 40, 120, and 160 mT. For clarity, 120 and 160 mT data are shifted down by 0.5 and 1.0 k$\Omega$ respectively. (b) $S_{V}^{1/2}/V$ (open circles) and $dR/d\theta/R$ (solid squares) at $B=40$ mT vs. field angle $\theta$. Inset: $S_{V}^{1/2}/V$ vs $dR/d\theta/R$ together with a linear fit (solid line). Certain points corresponding to the spikes are removed. (c) $S_{V}^{1/2}/V$ (open circles) and $dR/d\theta/R$ (solid squares) at $B=160$ mT vs. $\theta$. Inset: $S_{V}^{1/2}/V$ vs $dR/d\theta/R$ shows nonlinear relation.}\label{fig:angle_dependence}
\end{center}
\end{figure}

By varying the orientation of an in-plane magnetic field $B$ that satisfies $B_K<B<B_{AF}$ (where $B_K$ is the free layer anisotropy field) it is possible to smoothly rotate the magnetic moment of the free layer. Fig.~\ref{fig:angle_dependence} (a) shows the measured resistance as a function of $\theta$ (the in-plane angle between $B$ and nominal easy axis) for three values of $B$. The solid lines represent fits to the equation~\cite{sun_JMMM2008}
\begin{equation}
R(\varphi(\theta))=\frac{R_P R_{AP}}{R_P+(R_{AP}-R_P)\cos^2(\varphi(\theta)/2)},
\end{equation}

\noindent where $R_P$ and $R_{AP}$ are the resistance of the parallel and antiparallel state respectively, and $\varphi=\varphi_f-\varphi_{r}$, the angle difference between the free and reference layer magnetic moments, is a function of $\theta$. The functions $\varphi_f(\theta)$ and $\varphi_{r}(\theta)$ are found by minimizing the energies of the free and reference layers, which in the simplest approximation are given by~\cite{cullity_Wiley2008}
\begin{eqnarray}
E_f=-M_f V_f (B \cos(\theta-\varphi_f)+\frac{1}{2}B_K \cos^2\varphi_f)\\
E_{r}=-M_r V_r (B \cos(\theta-\varphi_{r})+B_{AF} \cos \varphi_{r}).
\end{eqnarray}

\noindent Here, $M_f$ ($M_r$) and $V_f$ ($V_r$) are the magnetization and volume of the free (reference) layer. The best fit with parameters $B_K=25$ mT and $B_{AF}=410$ mT accurately describes the device behavior at $B=120$ mT and $B=160$ mT but shows slight disagreement at $B=40$ mT, where the resistance displays sudden changes at particular field orientations, likely indicating that $B$ is not large enough to overcome pinning fields from defects.

We integrate the noise spectrum over the measured frequency range from 6 Hz to 200 kHz for each $\theta$. The average noise voltage as a function of field angle for $B=40$ mT ($B=160$ mT) is shown in Fig.~\ref{fig:angle_dependence} (b) [Fig.~\ref{fig:angle_dependence} (c)]. For $B=40$ mT,  $S_V^{1/2}(\theta)$ is very similar to $\chi_{MR}(\theta)$ obtained from Fig.~\ref{fig:angle_dependence} (a). Both curves show minima at $\theta$=0 and $\theta=\pi$ (P and AP states) and maxima close to $\theta=\pi/2$ and $\theta=3\pi/2$. Several spikes that appear in both the noise and $\chi_{MR}$ also indicate that excess noise is produced at field directions where sudden changes in the magnetic moment orientation occur. It is possible that irregular nanopillar edges, material defects, and grain boundaries provide pinning directions at which the magnetic moment is more difficult to rotate; a larger change in field orientation is required to unpin and rotate the magnetic moment resulting in sudden jumps in resistance. Just before depinning occurs, thermal excitations are able to excite the magnetic moment  between the pinned and the ``free'' orientations, causing enhanced noise with strongly Lorenzian behavior (data not shown here).

From the inset of Fig.~\ref{fig:angle_dependence} (b), a linear relation between $S_{V}^{1/2}/V$ and $dR/d\theta/R$ is phenomenologically established as
\begin{equation}
\frac{S_{V}^{1/2}}{V}=\frac{S_{V}^{1/2}(0)}{V}+k \left(\frac{1}{R} \frac{dR}{d\theta} \right),
\end{equation}

\noindent Here, $S_{V}^{1/2}(0)/V$=3.2$\times$10$^{-7}$ Hz$^{-1/2}$ represents the nonmagnetic background noise, while the calculated thermal and shot noise level is 2.5$\times$10$^{-7}$ Hz$^{-1/2}$ at 5 $\mu A$ and temperature of 10 K. The slope k of the fitting line in the inset of Fig.~\ref{fig:angle_dependence} (b) is equal to 1.6$\times$10$^{-6}$ rad/Hz$^{1/2}$. The linear relation established here between $S_{\theta}^{1/2}$ and $dR/d\theta/R$ is similar in spirit to the linear relation obtained by Ren $et$ $al.$~\cite{Ren_PRB2004} between the magnetic-field noise $S_{H}^{1/2}$ and $dR/dH/R$.

For larger fields, such as $B=160$ mT, [see Fig.~\ref{fig:angle_dependence} (c)], the noise minima at $\theta=\pi$ disappears, the similarity between the noise and $\chi_{MR}$ becomes less apparent (particularly around $\theta=\pi$), and the linear relation between noise and $\chi_{MR}$ does not hold [inset to Fig.~\ref{fig:angle_dependence} (c)]. As discussed in the previous section, when the field opposing the reference layer becomes comparable to $B_{AF}$, the free layer is no longer the dominant source of noise~\cite{ozbay_APL2009}. Although the fluctuations of the reference layer do not cause a clearly observable change in the resistance, they can dominate the spectrum at $\theta=\pi$: whereas the free layer is stabilized by the applied field, larger fields will continue to destabilize the SAF layers, providing additional mechanisms for fluctuations to occur.

\section{CONCLUSION}
In summary, we have measured low frequency noise of sub-micron MgO-based MTJs at 10 K. In most cases, the noise spectra are Lorentzian in character, indicating that noise is produced by a few magnetic fluctuators which are thermally excited in the free layer magnetic electrode. We experimentally established a linear relation between normalized noise $S_{V}^{1/2}/V$ and the angular magnetoresistance susceptibility $\chi_{MR}$ for $B\ll B_{AF}$, where noise is dominated by magnetization fluctuations of the free layer, and observed that for $B\sim B_{AF}$, destabilization of the reference layer due to SAF layer decoupling leads to a transition from free layer to reference layer dominated noise.

\begin{acknowledgments}
We wish to thank Jonathan Sun (IBM-MagIC MRAM Alliance, IBM T.J. Watson Research Center at Yorktown Heights, NY) for providing us with the devices and for helpful discussions. Part of this work was supported by the Army Research Office (Grant No. W911NF-08-0299).
\end{acknowledgments}


%
%

%



\end{document}